\documentclass[twocolumn,noshowpacs,amsmath,amssymb]{revtex4}
\usepackage{graphicx}
\usepackage{dcolumn}
\begin{document}
\title{ABC EFFECT AS A SIGNAL OF CHIRAL SYMMETRY RESTORATION \\ IN HADRONIC COLLISIONS}
\author{\firstname{M.~N.}~\surname{Platonova}}
\email{platonova@nucl-th.sinp.msu.ru}
\author{\firstname{V.~I.}~\surname{Kukulin}}
\email{kukulin@nucl-th.sinp.msu.ru} \affiliation{Skobeltsyn
Institute of Nuclear Physics, Lomonosov Moscow State University,
Moscow 119991, Russia}
%
%
\begin{abstract}
A new nonconventional mechanism for the basic $2\pi$-fusion
reaction $pn \rightarrow d+(\pi\pi)_0$ in the energy region $T_p =
1.0$--$1.4$~GeV is suggested. The mechanism is aimed at providing
a consistent explanation for the comprehensive experimental
studies of this reaction in exclusive setting done recently by the
WASA-at-COSY Collaboration. The basic assumption of the model
proposed is the production of the $I(J^P)=0(3^+)$ dibaryon
resonance $D_{03}$ in the $pn$ collision. The interference of two
decay channels of this resonance: $D_{03} \rightarrow d+\sigma
\rightarrow d + (\pi\pi)_0$ and $D_{03} \rightarrow D_{12} +\pi
\rightarrow d + (\pi\pi)_0$ is shown to give a strong
near-threshold enhancement in the $\pi\pi$ invariant mass
spectrum, which is well known as the ABC effect. The
$\sigma$-meson parameters found to reproduce the ABC enhancement
are in a general agreement with models which predict the chiral
symmetry restoration at high excitation energy and/or high density
of matter, although they are essentially less than those accepted
for the free $\sigma$ meson. So, this result might be considered
as an indication of partial chiral symmetry restoration in dense
and excited quark matter.
\end{abstract}
\maketitle
\newpage
The famous Abashian-Booth-Crowe (ABC) effect discovered more than
50 years ago~\cite{ABC} is observed in double-pionic fusion
reactions~\cite{ABC,Plouin,Banaigs} as a pronounced spectral
enhancement of isoscalar nature just above the $\pi\pi$-production
threshold. The effect was initially interpreted~\cite{ABC} as
being due to strong $\pi\pi$ rescattering in the scalar-isoscalar
channel, associated naturally with the $\sigma$ meson. However
later on the interpretation was left since no narrow resonance
with an appropriate mass ($m \simeq 300$~MeV) was found in
$\pi\pi$ scattering at low energies. At the same time, an other
interpretation~\cite{RS} for the ABC effect, based on a generation
of two $\Delta$ isobars via the $t$-channel meson exchange and
their subsequent decays with pion emission, was commonly accepted.
Although the ``$t$-channel $\Delta\Delta$'' mechanism did not
provide quantitative description of the data, it allowed to
reproduce the shape of differential cross sections found in the
numerous inclusive experiments on double-pionic
fusion~\cite{RS,Barry}.

The situation has changed dramatically quite recently, after
publication of the results of the first \emph{exclusive and
kinematically complete experiments} for the basic $2\pi$-fusion
reaction $pn \rightarrow d+\pi^0\pi^0$ done by the
CELSIUS/WASA~\cite{CW-EXP-0} and then by the WASA-at-COSY
Collaborations~\cite{CW-EXP}. The comparison of the new
experimental data with theoretical predictions has demonstrated
clearly that the above $t$-channel $\Delta\Delta$ model cannot
reproduce even the qualitative behaviour of the experimental
energy and angular distributions, giving just a low background in
the considered energy region ($T_p = 1.0$--$1.4$~GeV). At the same
time, the most intriguing discovery of these exclusive experiments
was an observation of a pronounced resonance structure in the
total $2\pi$-production cross section. This fact has been
interpreted as a generation of the dibaryon resonance $D_{03}$ in
the $pn$ collision, with quantum numbers $I(J^P) = 0(3^+)$, the
mass $m_{D_{03}}^{} \simeq 2.37$~GeV and the total width
$\Gamma_{D_{03}} \simeq 70$~MeV~\cite{CW-EXP}. Such a resonance
state has been predicted already in 1964 by Dyson and
Xuong~\cite{Dyson} and since then studied in numerous works, both
theoretical~\cite{Ping,Yuan,Wong,Valcarce} and
experimental~\cite{Ikeda}. From the new exclusive
experiments~\cite{CW-EXP}, the direct interrelation between the
production and decay of the $D_{03}$ resonance and the ABC effect
has been clearly established. Having considered the $D_{03}$ as
the $\Delta\Delta$ bound state, Adlarson \emph{et
al.}~\cite{CW-EXP} performed microscopic calculations based on the
mechanism $pn \rightarrow D_{03} \rightarrow \Delta\Delta
\rightarrow d + \pi^0\pi^0$. With such an ``$s$-channel
$\Delta\Delta$'' model they succeeded in a very good description
of the numerous energy and angular distributions observed in the
reaction $pn \rightarrow d+\pi^0\pi^0$. However, a reasonable
agreement with the experimental data at low $\pi\pi$ invariant
masses (in the region of the ABC peak) could be reached in their
work~\cite{CW-EXP} only when using a very soft form factor
$f_{\Delta\Delta}$ for the $D_{03} \rightarrow \Delta\Delta$
vertex with the cut-off parameter $\Lambda_{\Delta\Delta} =
0.15$~GeV/$c$. Such a low value of $\Lambda_{\Delta\Delta}$ means
the characteristic radius of the $D_{03}$ state to be even larger
than that of the deuteron. This is incompatible with the observed
strong $\Delta$--$\Delta$ binding in the $D_{03}$ state,
$\epsilon_{B}(D_{03}) \simeq 90$~MeV, and also with the results of
the various microscopic quark model calculations (see,
e.g.,~\cite{Yuan,Wong}), which all predict the radius for the
$0(3^+)$ $\Delta\Delta$ bound state $r(D_{03}) \simeq
0.7$--$0.9$~fm, i.e. of the order of the nucleon one. Hence, the
$D_{03}$ resonance appears to be the truly dibaryon state which
arises in a situation when the quark cores of two $\Delta$'s are
almost fully overlapped with each other. Moreover, the large width
of the free $\Delta$ isobar, $\Gamma_{\Delta} \simeq 120$~MeV,
would not allow for two $\Delta$'s to go away to far distance, so
the $D_{03}$ system, even after the pion emission, is likely to
stay in a dibaryon state with a small radius. So, this picture
contradicts essentially to the concept of the bound state of two
isolated quasi-free $\Delta$ isobars, which therefore looks to be
rather inconsistent. As will be shown below, a reasonable
explanation of the ABC effect in the basic $2\pi$-fusion reaction
may be found within an alternative model, involving the
$\sigma$-meson emission from the $D_{03}$ dibaryon and tightly
connected to the idea of chiral symmetry restoration in dense and
excited hadronic systems.

In constructing such a model, we start from the dibaryon concept
for short-range nuclear force~\cite{DBM1}. In this concept, the
conventional $t$-channel $\sigma$-meson exchange between two
isolated nucleons is replaced by the $s$-channel $\sigma$ exchange
with the $\sigma$ field surrounding the whole $6q$ bag, which
appears in the overlap region of two nucleons. An emission of a
light scalar meson occurs within a virtual transition of the $6q$
bag from the initial $2\hbar\omega$-excited quark configuration
$|s^4p^2[42]\rangle$ to its ground state $|s^6[6]\rangle$. Then, a
strong attraction of the $\sigma$ field to the multi-quark core
effectively induces a strong $NN$ attraction at intermediate
distances $r_{NN} \simeq 0.7$--$0.8$~fm. The predictions of the
model for the empirical $NN$-scattering phase shifts as well as
for the lightest nuclei properties are at the same level of
accuracy as those of other modern $NN$-force models, like Bonn and
Argonne, still keeping quite moderate values for short-range
cut-off parameters ($\Lambda_{\pi NN}$, etc.), which are
compatible with the QCD and quark model estimations (for the
details, see~\cite{DBM1,DBM2} and references therein to the
earlier works).

According to the dibaryon model, the deuteron wavefunction,
besides the conventional $NN$ component, has also a second,
quark-meson component, which becomes dominant at short $NN$
distances, i.e. when two nucleons are essentially overlapped with
each other~\cite{FN1}. The second component of the deuteron has
the structure $D_{01} \sim s^6 + \sigma \quad (l_{\sigma}=0,2)$ (a
compact $6q$ bag dressed with a $\sigma$ field), so it is similar
in some sense to the picture of the physical nucleon in which the
$3q$ core is dressed with a pionic cloud. Thus, analogously to the
excited states of the nucleon, one can examine the excited states
of the dibaryon $D_{01}$ and classify them on their total angular
momentum, isospin and parity. In this way, the experimentally
observed $D_{03}$ can be considered as a rotationally excited
state of the $D_{01}$, with the quark-meson structure $s^6 +
\sigma \quad (l_{\sigma}=2,4)$.

In fact, almost all dibaryon states lie in the vicinity of
two-baryon thresholds, e.g., $NN$, $N\Delta$, $\Delta\Delta$,
etc., and are coupled strongly to the respective two-baryon
channels. In our case, it is relevant to consider the following
chain of dibaryon states with rising angular momenta: $D_{01} \sim
NN$, $D_{12} \sim N\Delta$, $D_{03} \sim \Delta\Delta$, etc. Here
the $D_{12}$ is the isovector dibaryon resonance with quantum
numbers $I(J^P)=1(2^+)$ and the mass $m_{D_{12}}^{} \simeq
2.15$~GeV, discovered in the analysis of $pp$ scattering in the
${}^1D_2$ partial wave~\cite{Arndt1,Hosh1}. The production of the
$D_{12}$ resonance was later confirmed in $\pi^+d$ elastic
scattering~\cite{Arndt2,Hosh2} and particularly in the reaction
$\pi^+d \rightarrow pp$~\cite{Arndt3}, where the total cross
section at the energies $T_{\pi} \lesssim 200$~MeV is dominated by
the $D_{12}$-excitation process. Although the $D_{03}$ is a deeply
bound state in the $\Delta\Delta$ channel, it is a resonance in
the $p+n$ (as was observed in~\cite{CW-EXP}) and $D_{12}+\pi$
systems. It becomes a resonance also in the $d(D_{01})+\sigma$
system, if the $\sigma$ mass is less than $500$~MeV. So, there are
two basic possibilities for the decay of the $D_{03}$ resonance
into the deuteron (i.e. into its quark-meson component $D_{01}$)
and two pions:

$(i)$ by an emission of the $\sigma$ meson (mainly in the $d$ wave
relative to the $6q$ core due to the angular momentum
conservation) which then decays into two pions;

$(ii)$ by a sequential emission of two pions (each in the $p$
wave) through an intermediate isovector dibaryon $D_{12}$.

It is indicative that the above two interfering mechanisms for the
excited dibaryon decay $D_{03} \rightarrow d + \pi\pi$ can be
confronted with the quite similar two mechanisms for the Roper
resonance (excited nucleon) decay $N^*(1440) \rightarrow
N+\pi\pi$~\cite{PDG}: $N^*(1440) \rightarrow
N+(\pi\pi)_{I=0}^{s\text{-wave}} (\sigma)$ and $N^*(1440)
\rightarrow \Delta + \pi$. It should be stressed that the
model~\cite{Alv} based on an excitation of the Roper resonance and
its subsequent decay via these two channels was quite successfully
applied to the reactions $NN \rightarrow d+\pi\pi$ and $NN
\rightarrow NN+\pi\pi$ at the energies $T_N < 1$~GeV.

Thus, we consider the following resonance mechanisms related to
the above $D_{03}$ decay channels $(i)$ and $(ii)$ as the basic
contributions to the reaction $pn \rightarrow d+(\pi\pi)_0$ in the
ABC region ($T_p = 1.0$--$1.4$~GeV):

$(a)$ $pn \rightarrow D_{03} \rightarrow d+\sigma, \, \sigma
\rightarrow (\pi\pi)_0;$

$(b)$ $pn \rightarrow D_{03} \rightarrow D_{12}+\pi, \, D_{12}
\rightarrow d + \pi.$

\noindent The diagrams for these processes are shown in Fig.~1.

\begin{figure}[!ht]
\begin{center}
\resizebox{1.0\columnwidth}{!}{\includegraphics{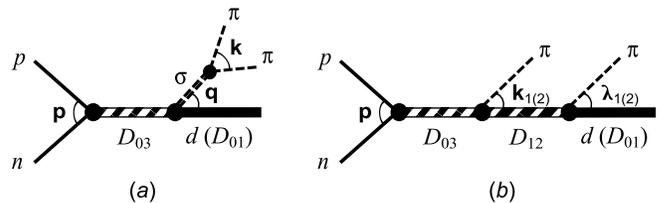}}
\end{center}
\caption{The leading mechanisms for the reaction $pn \rightarrow d
+ (\pi\pi)_0$ in the ABC region. The $3$-momenta in the c.m.s. of
two particles are indicated between the respective lines.}
\label{fig1}
\end{figure}

The amplitude for the emission of two neutral pions in the
reaction $pn \rightarrow d + \pi^0 \pi^0$ at the c.m. energy
$E=\sqrt{s}$ is then given by a sum of two terms:
\begin{equation}
\label{M} \mathcal{M}_{\mu_i\mu_f} =
\mathcal{M}^{(D_{03})}_{\mu_i}\left(\mathcal{M}^{(\sigma)}_{\mu_i\mu_f}+
\mathcal{M}^{(D_{12})}_{\mu_i\mu_f}\right),
\end{equation}
where
\begin{eqnarray}
\label{M03} \mathcal{M}^{(D_{03})}_{\mu_i} &=&
\frac{m_{D_{03}}^{2} \sqrt{\Gamma^{(2)}_{D_{03}
np}/p}}{E^2-m_{D_{03}}^2+i m_{D_{03}}
\Gamma_{D_{03}}} \mathcal{J}^{(D_{03})}_{\mu_i\mu_i}\bigl(\hat{p}\bigr), \\
\label{MS} \mathcal{M}^{(\sigma)}_{\mu_i\mu_f} \! &=& \!
\frac{m_{\sigma}\sqrt{\Gamma^{(2)}_{D_{03}d\sigma}/q} \, \sqrt{
\Gamma^{(0)}_{\sigma \pi^0 \pi^0}/k}}{M_{\pi\pi}^2-m_{\sigma}^2+i
m_{\sigma}\Gamma_{\sigma}} \mathcal{J}^{(\sigma)}_{\mu_i\mu_f}\bigl(\hat{q}\bigr), \\
\label{M12} \mathcal{M}^{(D_{12})}_{\mu_i\mu_f} \! &=& \!
\frac{1}{\sqrt{2}}\Biggl(\frac{m_{D_{12}}^{}\sqrt{\Gamma^{(1)}_{D_{03}
D_{12}\pi^0}/k_1} \, \sqrt{\Gamma^{(1)}_{D_{12} d
\pi^0}/\lambda_1}}{ M_{d\pi}^2-m_{D_{12}}^2+i m_{D_{12}}
\Gamma_{D_{12}}}
\nonumber \\
&\times& \mathcal{J}^{(D_{12})}_{\mu_i\mu_f}
\bigl(\hat{k}_1,\hat{\lambda}_1\bigr) +
\left[\vec{k}_1,\vec{\lambda}_1 \rightarrow
\vec{k}_2,\vec{\lambda}_2\right]\Biggr).
\end{eqnarray}
The amplitude $\mathcal{M}^{(D_{12})}_{\mu_i\mu_f}$ for the above
process $(b)$ is symmetrized over two identical pions~\cite{FN2}.

When taking into account only the dominating, i.e. the lowest,
partial waves in vertices (indicated in superscripts of $\Gamma$'s
in Eqs.~(\ref{M03})--(\ref{M12})), the spin-angular terms
$\mathcal{J}^{(\sigma)}_{\mu_i\mu_f}$ and
$\mathcal{J}^{(D_{12})}_{\mu_i\mu_f}$ can be calculated by using
the standard technique for the angular momenta coupling. Thus, the
total angular momentum $J$ should be decomposed as $J=J_1+L$, i.e.
$3=1+2$ for the process $(a)$ and $\{3=2+1, 2=1+1\}$ for the
process $(b)$. The factor
$\mathcal{J}^{(D_{03})}_{\mu_i\mu_i}(\hat{p})$ comes from the
vertex $np \rightarrow D_{03}$ and, with the initial momentum
$\vec{p}$ directed along $z$ axis, gives just a constant
$C_{\mu_i}$.

With the amplitudes defined in Eqs.~(\ref{M})--(\ref{M12}), the
differential cross sections as functions of the invariant masses
squared $M_{\pi\pi}^2$ and $M_{d\pi}^2$ are given by
\begin{eqnarray}
\label{DSPP} \frac{d\sigma}{d\left(M_{\pi\pi}^2\right)} \!\! &=&
\!\! \frac{\rho^{(\pi\pi)}}{(4\pi)^{5} p E}\! \int{\!\! \int{\!
d\Omega_q d\Omega_k
\frac{1}{3}\!\sum\limits_{\mu_i,\mu_f}{\! |\mathcal{M}_{\mu_i\mu_f}|^2}}}, \\
\label{DSDD} \frac{d\sigma}{d\left(M_{d\pi}^2\right)} \!\! &=&
\!\! \frac{\rho^{(d\pi)}}{(4\pi)^{5} p E} \! \int{\!\! \int{ \!
d\Omega_{k_1} \! d\Omega_{\lambda_1}
\frac{1}{3}\!\sum\limits_{\mu_i,\mu_f}{\!
|\mathcal{M}_{\mu_i\mu_f}|^2}}},
\end{eqnarray}
where $\rho^{(\pi\pi)} = qk/2E M_{\pi\pi}$ and $\rho^{(d\pi)} =
k_1 \lambda_1/2E M_{d\pi}$ are the Lorentz-invariant phase-space
factors. The sum should be taken over all possible projections
$\mu_i$ and $\mu_f$ of the total spin $S=1$ in initial and final
states, since the production of the dibaryon resonance with
quantum numbers $I(J^P)=0(3^+)$ can occur in the $np$ triplet
state only.

The energy dependence for the partial width of the resonance $R$
with the invariant mass $M$ decaying into particles $1$ and $2$
with invariant masses $M_1$ and $M_2$ and the relative orbital
angular momentum $l$ has been parameterized as
\begin{equation}
\label{G} \Gamma^{(l)}_{R12}(q) = \Gamma^{(l)*}_{R12}
\left(\frac{q}{q^*}\right)^{2l+1} \left(\frac{(q^*)^2 +
\varkappa^2}{q^2+\varkappa^2}\right)^{l+1},
\end{equation}
where $q = \left[(M^2-M_1^2-M_2^2)^2-4 M_1^2M_2^2\right]^{1/2}/2M$
is the modulus of the relative momentum between particles $1$ and
$2$, and an asterisk denotes the values in the resonance point.
Such a parametrization provides a correct near-threshold behaviour
of the partial widths, however preventing an unphysical rise of
the widths at higher energies (see~\cite{Huber} for a similar
parametrization in case $l=1$). Thus, with an appropriate value of
the parameter $\varkappa$, the centre of the Breit--Wigner
distribution can be properly reproduced. For the partial widths
introduced in Eqs.~(\ref{MS})--(\ref{M12}), this is achieved with
$\varkappa = 0.1$--$0.2$~GeV/$c$, while for the partial width
$\Gamma^{(2)}_{D_{03} np}$ entering Eq.~(\ref{M03}) one should use
the larger value $\varkappa = 0.35$~GeV/$c$.

The masses and total widths of the dibaryon resonances $D_{03}$
and $D_{12}$ have been fixed in our calculations as~\cite{FN3}
\begin{center}
$m_{D_{03}}=2370$~MeV, $\Gamma_{D_{03}}=\,70\,$~MeV, \\
$m_{D_{12}}=2150$~MeV, $\Gamma_{D_{12}}=110$~MeV.
\end{center}
The remaining model parameters, i.e. the mass and width of the
$\sigma$ meson and the relative weight of the amplitudes
corresponding to the processes $(a)$ and $(b)$, were derived from
the fit to the experimental data~\cite{CW-EXP} on the
$M_{\pi\pi}^2$ spectrum, and then the $M_{d \pi}^2$ spectrum was
calculated using the same parameter values.

The results for the $M_{\pi\pi}^2$ and $M_{d\pi}^2$ distributions
at the peak energy (where the total cross section has a maximum)
$\sqrt{s} = 2.38$~GeV, or $T_p = 1.14$~GeV, are presented in
Fig.~2. The results are normalised to the experimental value of
the total cross section $(\sigma_T)_{\rm peak} \simeq
0.43$~mb~\cite{CW-EXP}. It is evident that our simple model
reproduces the shapes of these two distributions almost perfectly.
We observe that, although the cross section for the
$\sigma$-generation process $(a)$ alone is rather moderate, its
contribution is crucial to reproduce the shape of the
$M_{\pi\pi}^2$ distribution. Thus, its constructive interference
with the $D_{12}$-production mechanism ($b$) at low $M_{\pi\pi}^2$
leads just to the observed height of the ABC peak. On the other
hand, when considering the $M_{d\pi}^2$ distribution (see
Fig.~2$b$), the $\sigma$-production mechanism plays a role of a
smooth background, while the process $(b)$ alone almost gives the
observed resonance enhancement.

The resonance peak in the $M_{d\pi}^2$ spectrum was associated
previously with an excitation of the intermediate $\Delta$
isobar~\cite{CW-EXP}. However, our results show that this peak may
reflect just a generation of the intermediate isovector dibaryon
$D_{12}$ which then decays into the final deuteron and pion. In
fact, the suggested mechanism of two-pion emission through the
$D_{12}$ excitation is quite similar to the $s$-channel
$\Delta\Delta$ model used in~\cite{CW-EXP}, however without a soft
form factor $f_{\Delta\Delta}$. The point is that the $D_{12}$
dibaryon is located near to the $N\Delta$ threshold, so it has a
large probability of being in the $N+\Delta$ (${}^5S_2$) state.
Therefore, one needs additional tests to distinguish between these
two mechanisms.

We also calculated the angular distributions for the final
deuteron and pion emissions in the overall c.m. frame and then
compared our model predictions with the experimental data. Our
results for the angular distributions are shown in Fig.~3. The
agreement with the data is not as good as for the invariant mass
distributions, however it is still quite reasonable. Moreover, if
we confront our model predictions with those found
in~\cite{CW-EXP} on the basis of the $s$-channel $\Delta\Delta$
model, the description of the above two angular distributions
seems to be not worse than that reached in~\cite{CW-EXP}. So, with
only three basic parameters extracted from the experimental
$M_{\pi\pi}^2$ spectrum, our simple model is able to reproduce
four differential distributions measured in the $pn \rightarrow
d+\pi^0\pi^0$ reaction.

\begin{figure}[ht]
\begin{center}
\resizebox{1.0\columnwidth}{!}{\includegraphics{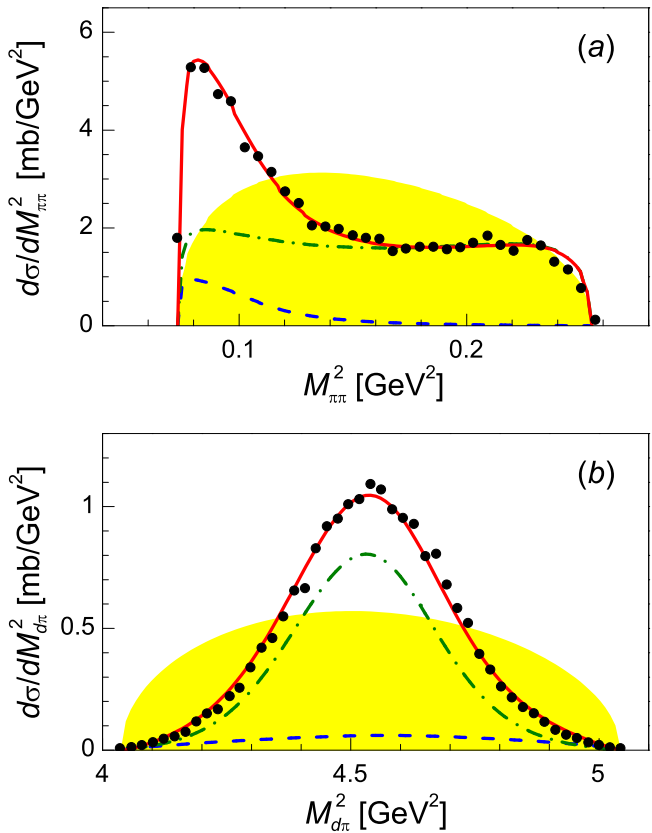}}
\end{center}
\caption{ (Color online) Differential cross sections as functions
of the invariant masses squared $(a)$ $M_{\pi\pi}^2$ and $(b)$
$M_{d\pi}^2$ in the reaction $pn \rightarrow d+\pi^0\pi^0$ at the
energy $\sqrt{s} = 2.38$~GeV. The contribution of the
$\sigma$-production mechanism (see Fig.~1$a$) is shown by dashed
lines while the contribution of the mechanism going through the
intermediate dibaryon $D_{12}$ (see Fig.~1$b$) is shown by
dash-dotted lines. The solid lines correspond to the summed cross
sections. Shaded areas show the pure phase-space distributions.
The experimental data (full circles) are taken from
Ref.~\cite{CW-EXP}.} \label{fig2}
\end{figure}

\begin{figure}[ht]
\begin{center}
\resizebox{1.0\columnwidth}{!}{\includegraphics{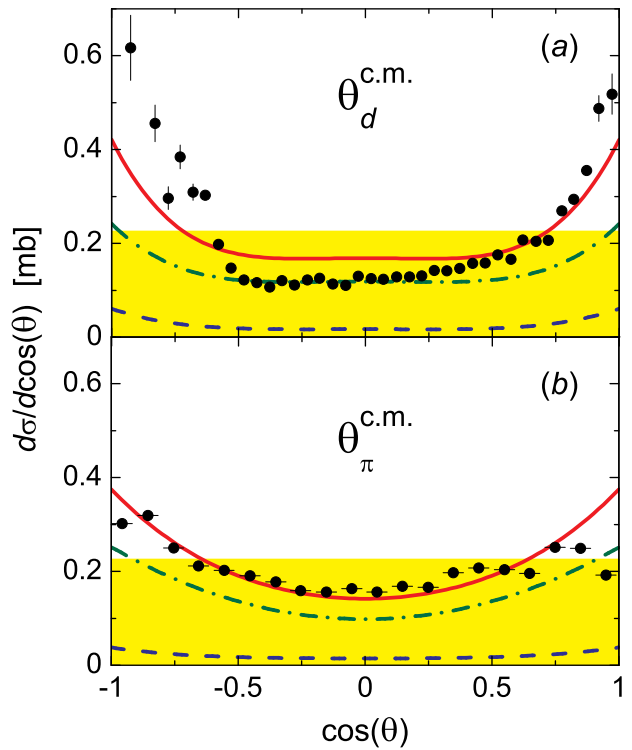}}
\end{center}
\caption{ (Color online) Angular distributions for the deuteron
$(a)$ and the pion $(b)$ in the overall c.m.s. at the energy
$\sqrt{s} = 2.38$~GeV. The meaning of curves is the same as in
Fig.~2. The experimental data (full circles) are taken from
Ref.~\cite{CW-EXP}.} \label{fig3}
\end{figure}

Besides the two considered decay modes of the $D_{03}$ resonance,
one can also treat other channels for the decay $D_{03}
\rightarrow d+\pi\pi$, i.e. via a simultaneous emission of two
uncorrelated pions without formation of the $\sigma$ meson or a
sequential emission of two pions through other intermediate
isovector dibaryons~\cite{Arndt1,Hosh1}, such as the $1(3^-)$
state (corresponding to the ${}^3F_3$ $NN$ partial wave). In these
cases, the pions may be emitted in $s$- and $d$-waves relative to
the $6q$ bag, thus forming the $d$-wave $\pi\pi$ pair. In fact,
one can see from Fig.~3 that the deuteron and pion c.m. angular
distributions show some additional $d$-wave admixture which is not
fully taken into account by the present model. Including the
corrections from the $d$-wave pion emission would not affect
significantly the shape of the invariant mass spectra, but may
improve the description of the angular distributions at forward
and backward directions. In a complete theoretical picture the
conventional $t$-channel $\Delta\Delta$ mechanism should also be
taken into account as the main background process to the $D_{03}$
production.

The mass and width of the $\sigma$ meson extracted from the fit to
the ABC peak are
\begin{center}
$m_{\sigma} \simeq 300$~MeV, $\Gamma_{\sigma} \simeq 100$~MeV.
\end{center}
These values are notably less than those for the free $\sigma$
mass and width, found by extrapolation from the dispersion
relations for the $\pi\pi$ scattering amplitude to the $\sigma$
complex pole~\cite{Caprini},
\begin{center}
$m_{\sigma}^{(0)}=441^{+16}_{-8}$~MeV,
$\Gamma_{\sigma}^{(0)}=544^{+18}_{-25}$~MeV.
\end{center}
While the latter values are within the range for the $f_0(500)$ or
$\sigma$ pole positions currently quoted in PDG tables~\cite{PDG}:
\begin{center}
$m_{\sigma}=400$--$550$~MeV, $\Gamma_{\sigma}=400$--$700$~MeV,
\end{center}
the values found here are essentially out of this range. To
resolve this discrepancy, one should bear in mind that the above
range of pole positions for the $\sigma$ meson was fixed by
including only those analyses consistent with the low-energy
$\pi\pi$ scattering data as well as the advanced dispersion
analyses such as performed in~\cite{Caprini}. On the other hand,
numerous theoretical investigations (see, e.g.,~\cite{HK,Volkov})
show that the mass and width of the $\sigma$ meson produced in hot
and/or dense nuclear matter may be significantly shifted downwards
due to the partial chiral symmetry restoration (CSR) effect.
Besides that, it was demonstrated~\cite{Glozman} that the partial
CSR takes place also in strongly excited states of \emph{isolated
hadrons} (baryons and mesons) at the excitation energies $E^*
\gtrsim 500$~MeV. In particular, the appearance of approximately
degenerate parity doublets in the spectra of highly excited
baryons may be considered as a direct manifestation of partial
CSR. In fact, the rise of baryon density or nuclear matter
temperature as well as a high hadron excitation energy leads to an
increase of quark kinetic energy, which results in the suppression
of the chiral condensate in QCD vacuum. This, in turn, means the
reduction of the $\sigma$-meson mass and the width for the $\sigma
\rightarrow \pi\pi$ decay. So, the $\sigma$ meson, being a broad
resonance in free space, may become a sharp resonance in dense or
excited hadronic media.

We emphasize that within the dibaryon model~\cite{DBM1,DBM2}, the
best description of the $NN$-scattering phase shifts and the
properties of the lightest nuclei has been achieved with a rather
low mass of the $\sigma$ meson, $m_{\sigma} \simeq 350$~MeV,
whereas in the conventional meson-exchange $NN$-force models the
$\sigma$ mass is taken to be $500$--$600$~MeV. Since in the
dibaryon model the initial $6q$ bag (with the quark configuration
$|s^4p^2[42]\rangle$) is a dense object ($r_{6q} \simeq
0.5$--$0.6$~fm) and is also the $2\hbar\omega$-excited hadronic
state, the renormalisation of the $\sigma$ mass in the field of
the bag might be related to the partial CSR~\cite{DBM1}. The
situation is quite similar for the $D_{03}$ resonance, which also
represents dense quark matter (the density of a $6q$ system with a
radius $r \simeq 0.8$~fm corresponds to about six-fold normal
nuclear density) and has an additional excitation energy of
$500$~MeV above the deuteron pole. Thus, the $\sigma$ meson
produced from the $D_{03}$ decay should have the lower mass and
width than those for the free $\sigma$ meson. As the $\sigma$
width found here is still quite large, the $\sigma$ meson is
likely to decay \emph{before} it escapes the field of the
multi-quark bag and acquires its free-space parameters. This
implies that when measuring the $\pi\pi$ invariant mass
distribution, one should observe just the renormalised $\sigma$
meson with the reduced mass and width. So, one can suggest that
the low values for the $\sigma$-meson parameters found here
indicate a partial CSR in the excited dibaryon state. This
conclusion is in an agreement with the results of numerous
theoretical studies concerning the CSR in hadronic and nuclear
media~\cite{HK,Volkov,Glozman}. Further experimental and
theoretical efforts are called for to check the fundamental CSR
effects in hadronic systems.

\emph{To summarize}, we have proposed a new nonconventional model
for the basic double-pionic fusion reaction $pn \rightarrow
d+(\pi\pi)_0$ in the ABC region ($T_p = 1.0$--$1.4$~GeV). The
model takes into account the $D_{03}$-dibaryon production and its
decay into the final deuteron and two pions by two alternative
ways: $(i)$ through an emission of the $\sigma$ meson and $(ii)$
through a generation of the intermediate isovector dibaryon
resonance $D_{12}$. So, the suggested mechanisms for the $D_{03}$
decay have a remarkable reminiscence with two analogous modes of
the Roper resonance $N^*(1440)$ decay. A reasonable agreement with
the data of the recent exclusive experiments done by the
WASA-at-COSY Collaboration~\cite{CW-EXP}, \emph{without} an
assumption of the unnaturally soft form factor in the vertex
$D_{03} \rightarrow \Delta\Delta$, is obtained.

Within the model proposed, the ABC effect is considered as a
result of the $\sigma$-meson emission, whose mass and width, due
to the partial restoration of chiral symmetry, are reduced in the
field of the multi-quark bag as compared to their free-space
values. In this way, the observed enhancement in the
low-$M_{\pi\pi}$ spectrum, similarly to the instant photograph,
shows just the renormalised $\sigma$ meson in the field of the
bag. Hence, by extracting the $\sigma$ mass and width from the
experimentally measured ABC peak, one is able to judge the degree
of chiral symmetry restoration in excited and/or dense hadronic
systems. With this interpretation, it is easily understood why the
low-$M_{\pi\pi}$ enhancement is not seen in the reaction $pn
\rightarrow pp + \pi^- \pi^0$~\cite{Skorodko}: although the
$D_{03}$ resonance is produced there as well, but \emph{the
$\sigma$ meson is not}. So, we partially rehabilitate the initial
interpretation of the ABC effect suggested by Abashian, Booth and
Crowe~\cite{ABC}, even though the $\sigma$-meson generation in our
model is not related to the $\pi\pi$ final-state interaction.
Thus, on the basis of the model proposed, one can treat ABC-type
experiments as a \emph{direct observation of the $\sigma$-meson
production} in $NN$, $Nd$, etc., collisions.

The authors are grateful to Prof.~H.~Clement, Drs.~M.~Bashkanov
and T.~Skorodko from Physical Institute of Tuebingen University
for fruitful discussions on the WASA-at-COSY experimental results.
One of the authors (V.I.K.) also thanks Prof.~A.~Faessler for the
hospitality at Tuebingen University where this work was begun. The
work was done under partial financial support from RFBR grants
Nos.~10-02-00096 and 12-02-00908.


\begin{thebibliography}{99}
\bibitem{FN1}
The quark-meson component gives a small contribution (ca.
$2$--$3$\%) to the total deuteron wavefunction normalisation, so
it should be visible only when probing the deuteron structure with
high-momentum probes~\cite{DBM1}.
\bibitem{FN3} The $D_{03}$ mass and width are taken
from Ref.~\cite{CW-EXP}, and the parameters chosen for the
$D_{12}$ are close to those found in Ref.~\cite{Hosh2}.
\end{thebibliography}
\end{document}